# Large Brightness Variations of Uranus at Red and Near-IR Wavelengths

Richard W. Schmude Jr.

Gordon State College

419 College Dr., Barnesville, GA 30204 USA

schmude@gordonstate.edu

Ronald E. Baker

Indian Hill Observatory, Chagrin Valley Astronomical Society

PO Box 11, Chagrin Falls, OH 44022 USA

rbaker52@gmail.com

Jim Fox

Astronomical League and the AAVSO

P.O. Box 135, Mayhill, NM 88339

Makalii45@gmail.com

Bruce A. Krobusek

5950 King Hill Road, Farmington, NY, 14425, USA

bkrobusek@gmail.com

Anthony Mallama

14012 Lancaster Lane, Bowie, MD 20715 USA

anthony.mallama@gmail.com

2015 October 14

Pages, 30; tables, 10; figures, 13

Abstract

Uranus is fainter when the Sun and Earth are near its equatorial plane than when they are near the projection of its poles. The average of the absolute values of the sub-Earth and sub-Sun latitudes (referred to as the sub-latitude here) is used to quantify this dependency. The rates of change of magnitude with sub-latitude for four of the Johnson-Cousins band-passes are B-band, -0.48 +/- 0.11 milli-magnitudes per degree; V-band, -0.84 +/- 0.04 ; R-band, -5.33 +/- 0.30; and I-band -2.79 +/- 0.41. Evaluated over the range of observed sub-latitudes, the blue flux changes by a modest 3% while the red flux varies by a much more substantial 30%. These disk-integrated variations are consistent with the published brightness characteristics of the North and South Polar Regions, with the latitudinal distribution of methane and with a planetary hemispheric asymmetry. Reference magnitudes and colors are also reported along with geometric albedos for the seven Johnson-Cousins band-passes.

1. Introduction

Uranus is the only planet in our solar system whose rotational axis is highly inclined to the ecliptic pole. This unique geometry permits favorable viewing of both high and low latitudes from the Earth. The passage of the Sun and Earth through the planet's equatorial plane in 2007 was observed closely. Unexpected storm activity (de Pater et al., 2015) and other atmospheric changes have been noted since the arrival of spring in the northern hemisphere.

At the same time, the planet has grown considerably brighter especially at red and near-IR wavelengths. In this paper we report on this post-equinox brightening and we analyze the historical record of standardized photometry extending back to the middle of the last century. The brightness changes are discussed in the context of disk-resolved Uranian imagery (Karkoschka, 2001b), spectrographic observations (Sromovsky et al., 2014) and characterization of the south polar region (Rages et al., 2004). This work is closely related to that of Lockwood and Jerzykiewicz (2006) who have recorded the brightness of Uranus in the b- and y-bands since 1972.

Section 2 summarizes a long series of b- and y-band photometry recorded by Lockwood and Jerzykiewicz and explains its relationship to our B- and V-band results. Section 3 provides an overview of the photometric and calibration techniques used to determine the magnitudes analyzed in this paper. Section 4 describes the small brightness changes that we and previous investigators observed in the B- and V-bands over 73% of a Uranian year and compares them to the expected change due to the planet's oblate figure. Section 5 describes the large brightness changes that were recorded in the R- and I-bands, which are being reported here for the first time. Section 6 examines the effects that the satellites, rings and clouds of Uranus might have on observed brightness. Section 7 discusses the variable planetary brightness in terms of the polar regions, methane depletion and hemispheric asymmetry. Section 8 lists reference magnitudes, colors and albedos. Section 9 summarizes our conclusions.

2. Related photometry in blue and yellow band-passes

The main subject of this paper is wide-band photometry of Uranus in visible and near-IR wavelengths. Before commencing on that discussion though, we summarize the important medium-band blue and yellow photometry of Lockwood and Jerzykiewicz (2006). This work has been updated on Lockwood's web page at [http://www2.lowell.edu/users/wes/U_N_lcurves.pdf](http://www2.lowell.edu/users/wes/U_N_lcurves.pdf) and now spans 42 years. The annual means of their b (472 nm) and y (551 nm) are plotted in Figs. 1 and 2 and they reveal quasi-sinusoidal variations peaking near the Uranian solstice of the 1980s. Lockwood notes that "The b amplitude slightly exceeds the expected 0.025 mag purely geometrical variation caused by oblateness as the planetary aspect changes, seen from Earth. The y amplitude is several times larger, indicating a strong equator-to-pole albedo gradient." This result anticipates the blue and green wide-band results which are presented in the present paper.

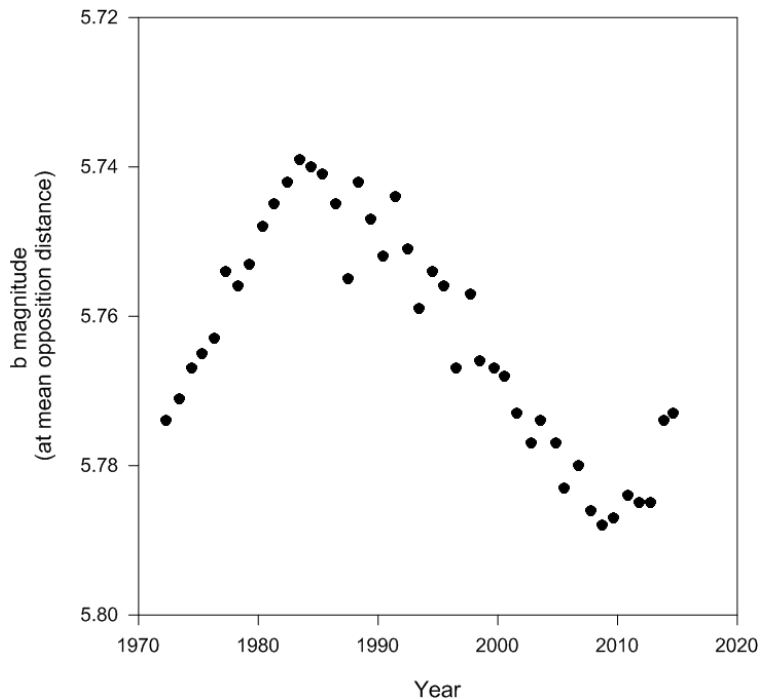

Fig. 1. The b-band variability of Uranus.

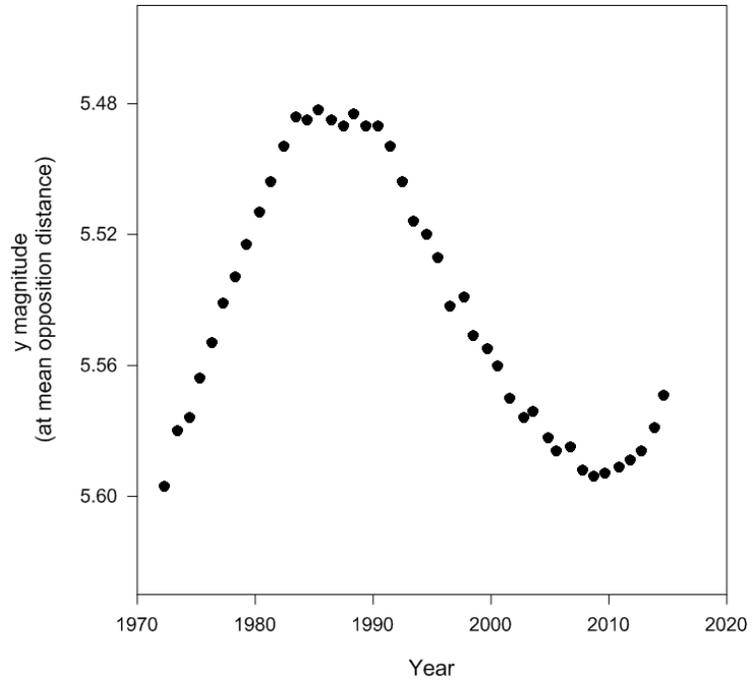

Fig. 2. The y-band variability of Uranus.

3. Photometry and calibrations

The magnitudes used in this study are on the Johnson-Cousins system. The original Johnson system was defined by Johnson et al. (1966) and the Cousins extension is described in Cousins (1976a and 1976b). The characteristics of the system are given in Table 1.

Table 1: Characteristics of the Johnson-Cousins systems

| Filter | Effective Wavelength (μm) | FWHM (μm) |
| --- | --- | --- |
| U  | 0.360 | 0.068 |
| B  | 0.436 | 0.098 |
| V  | 0.549 | 0.086 |
| R  | 0.700 | 0.209 |
| Rc | 0.641 | 0.158 |
| I  | 0.900 | 0.221 |
| Ic | 0.798 | 0.154 |

In order to sample the brightness of Uranus over a wide range of sub-Earth and sub-Sun latitudes we retrieved measurements dating back to the 1950s on the Johnson-Cousins magnitude system. The data sources are summarized in Table 2 and are described below.

Serkowski (1961) reports two sets of V and B-V magnitudes of Uranus made between October 30, 1953 and April 12, 1961 in his Table VIII. These values are corrected for extinction and color transformation (those corrections are described later in this section) but are not corrected for distance, phase and oblateness effects. One set is based on his system of primary standards and the other is based on his system of 10-year standards. The two sets of measurements have mean differences of 0.0025 and 0.001 magnitudes for the V and B-V sets. These are small and are assumed to be negligible. The data set based on primary standards is selected for this study because it contains more measurements. Each B value is the sum of the V and B-V value.

Jerzykiewicz and Serkowski (1966) reports V, B-V and B' magnitudes in their Table VI. We did not use their B' magnitudes because they are corrected for oblateness. The V and B-V values are used in this study. The B-V values are computed in the same way as in the Serkowski data set.

As a check on both studies, one of us (RWS) compared the measured V and B-V values of the Ten-Year Standards in Table VI of Serkowski (1961). Most of the values are consistent with those in Iriarte et al. (1965). Jerzykiewicz and Serkowski (1966) report that the Ten-Year Standard star Xi-Boötis varied by over 0.03 magnitudes in 1964. Sinnott and Perryman (1997) list this star as a variable changing by less than 0.1 magnitude. All of the other Ten-Year Standards are listed as having a constant brightness by Sinnott and Perryman (1997). This leads us to believe that the work of Serkowski (1961) and of Jerzykiewicz and Serkowski (1966) are of high quality.

Schmude (1992, 1993, 1994, 1996, 1997a, 1997b, 1998a, 1998b, 2000a, 2000b, 2001, 2002, 2004, 2005, 2006a, 2006b, 2008, 2009, 2010a, 2010b, 2012, 2013, 2014, 2015a and 2015b) and co-workers report V filter results for Uranus. In many cases, B, R and I results are also reported. All values are corrected for extinction. Data in 1997 and from 1999 and later were corrected for color transformation in the original reports. Therefore, no further modification was made to these data. The values collected from 1991-1995 and 1997 in Schmude (1992, 1993, 1994, 1996, 1997a, b, 1998b) were not corrected for color transformation in the original reports. Therefore, corrections were made before adding these measurements to the data set used in this study. The transformation coefficients were computed from the two-star method outlined in Hall and Genet (1988, p. 186). Between 1991 and 1999, the same photometer was used. In cases where there was no transformation data available, the data was not included in the current study. Starting in 1999, RWS used a new photometer and thus re-measured the transformation coefficients (Schmude, 2000b). Others besides RWS and JF contributed brightness measurements to this data set. Their names are listed in the original sources along with their transformation coefficients. The complete data set is available upon request from one of the authors (RWS).

New photometry acquired by two of the authors (BAK and REB) is summarized in Table 2 and the observations themselves are listed in Tables 3 and 4. Uncertainties for the observations by BAK and REB are listed with their observations. Formal error bars are not available for the other observations but they are estimated to be a few hundredths of a magnitude or less.

Table 2: Data sets used in the development of the photometric model of Uranus

```
  Years         Filters       Source
1953-1961       B V           Serkowski (1961)
1962-1966       B V           Jerzykiewicz and Serkowski (1966)
1991-2015       B V R I       Schmude (1992, 1993, 1994, 1996, 1997a,
                              b, 1998a, b, 2000a, b, 2001, 2002, 2004,
                              2005, 2006a, b, 2008, 2009, 2010a, b,
                              2012, 2013, 2014 and 2015a, b)
2013-2014       B V R I       Krobusek (this paper)
2014            B V Rc Ic     Baker (this paper)
```

Table 3: Magnitudes recorded by BAK

```
   MJD*        B     +/-      V     +/-      R     +/-      I     +/-
  56626.02  -6.605 0.014   -7.135 0.014   -6.824 0.022   -5.663 0.046
  56626.08  -6.598 0.020   -7.140 0.019   -6.897 0.020   -5.675 0.047
  56643.98  -6.629 0.013   -7.127 0.013   -6.915 0.032   -5.700 0.089
  56888.32  -6.613 0.017   -7.128 0.016   -6.920 0.032   -5.696 0.089
  56894.29  -6.625 0.017   -7.183 0.019   -6.947 0.035   -5.706 0.091
  56895.28  -6.609 0.030   -7.178 0.016   -6.918 0.035   -5.770 0.085
```
*Modified Julian Date
 "+/-" refers to uncertainty

Table 4: Magnitudes recorded by REB

| MJD*     | B      | +/-   | V      | +/-   | Rc     | +/-   | Ic     | +/-   |
|----------|--------|-------|--------|-------|--------|-------|--------|-------|
| 56876.37 | -6.620 | 0.008 | -7.113 | 0.008 | -6.987 | 0.022 | -6.085 | 0.047 |
| 56876.38 | -6.594 | 0.006 | -7.112 | 0.007 | -6.985 | 0.023 | -6.073 | 0.048 |
| 56924.26 | -6.618 | 0.006 | -7.134 | 0.008 | -6.991 | 0.023 | -6.085 | 0.048 |
| 56924.27 | -6.627 | 0.007 | -7.127 | 0.007 | -7.016 | 0.023 | -6.077 | 0.049 |
| 56942.23 | -6.623 | 0.006 | -7.128 | 0.007 | -6.995 | 0.022 | -6.086 | 0.048 |
| 56942.24 | -6.630 | 0.009 | -7.137 | 0.010 | -7.002 | 0.023 | -6.076 | 0.048 |

*Modified Julian Date

"+/-" refers to uncertainty

The observations were made differentially with respect to standard stars of the Johnson-Cousins system. The procedures for differential photometry were described by Hardie (1962) and again by Hall and Genet (1988).

All magnitudes were corrected for extinction and color effects. The extinction correction accounts for dimming due to the Earth's atmosphere by applying a correctional coefficient to the air mass through which the object was viewed. Color correction takes the 'natural' magnitudes recorded by the hardware and converts them to standard magnitudes on the Johnson-Cousins system. Equations 1 and 2 illustrate extinction correction and color transformation, respectively.

$$m_0 = m - k_M X \tag{1}$$

$$M = m_0 + \varepsilon C \tag{2}$$

In equation 1, *m* is the raw magnitude, $k_M$ is the extinction coefficient, *X* is the air mass and $m_0$ is the magnitude corrected for extinction. In equation 2, *ε* is the transformation coefficient, *C* is the color index and *M* is the magnitude corrected for extinction and color. More information on photometric correction can be found in Hall and Genet (1988) and in Hardie (1962).

A concern with regard to color transformation is that color coefficients are based on stellar spectra. However, planetary spectra are different and that of Uranus, in particular, is quite steep at long wavelengths as seen in Fig. 3.

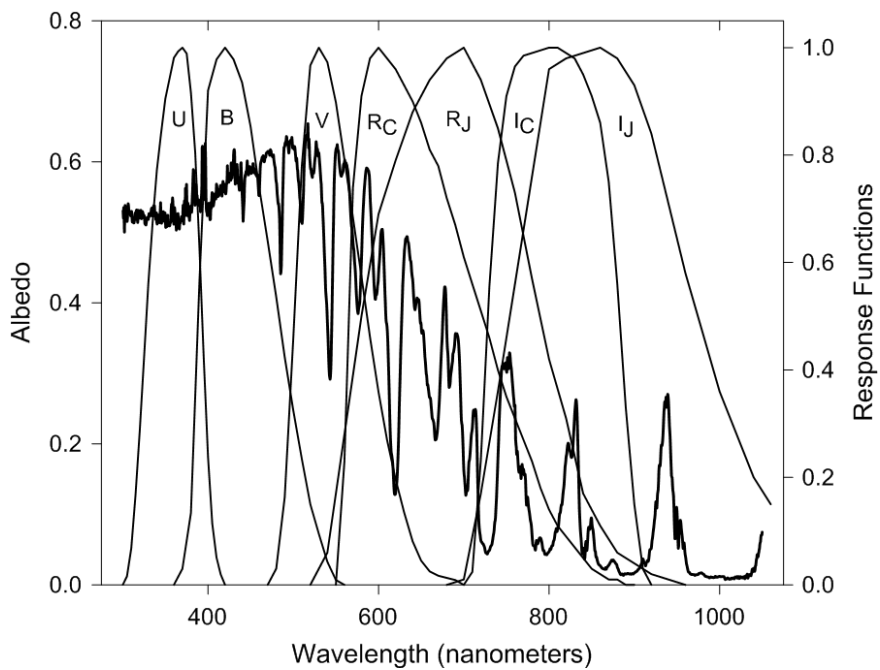

Fig. 3. The response functions of the Johnson-Cousins system (Filter Profile Service at http://skyservice.pha.jhu.edu) are superposed on the albedo of Uranus (Karkoschka, 1998).

After extinction correction and color transformation, the magnitudes were reduced to a standard distance of one astronomical unit from the Earth to Uranus and from the Sun to Uranus.

The phase angle of Uranus can only be as large as three degrees, so its effect on the planet's magnitude is expected to be very small. In order to estimate the size, we evaluated the phase functions of Jupiter in U, B, V, R and I (Mallama and Schmude, 2012) and of Venus in B, V, R and (Mallama et al., 2006) at the three degree limit for Uranus. In all cases the magnitude change is 0.01 magnitude or less. Likewise, Lockwood and Thompson (2009) found that the phase function for Titan is well under 0.01 magnitude per degree. These results are all consistent with results reported by Lockwood (1978) that the effect of solar phase angle is less than 0.003 magnitude. Thus, the effect of solar phase angle on the brightness of Uranus is not significant and, since the exact phase function is not known, no correction was applied. Likewise, no correction was made for oblateness but that effect will be discussed later in the paper. Finally, to conclude this section, mean magnitudes for each opposition are listed in Table 5.

Table 5: Mean magnitude and number of observations for each opposition

```
 Year        B     #     V     #     R     #     I     #    Rc     #    Ic     #
1954.04   -6.583  11  -7.126  11   ----   0   ----   0   ----   0   ----   0
1955.05   -6.632   6  -7.165   6   ----   0   ----   0   ----   0   ----   0
1956.06   -6.621  18  -7.150  18   ----   0   ----   0   ----   0   ----   0
1957.07   -6.611  12  -7.133  12   ----   0   ----   0   ----   0   ----   0
1958.08    ----    0   ----    0   ----   0   ----   0   ----   0   ----   0
1959.10   -6.624  10  -7.146  10   ----   0   ----   0   ----   0   ----   0
1960.11   -6.601  17  -7.130  17   ----   0   ----   0   ----   0   ----   0
1961.12   -6.605   8  -7.131   8   ----   0   ----   0   ----   0   ----   0
1962.13   -6.597  15  -7.120  15   ----   0   ----   0   ----   0   ----   0
1963.14    ----    0   ----    0   ----   0   ----   0   ----   0   ----   0
1964.16   -6.587   2  -7.095   2   ----   0   ----   0   ----   0   ----   0
1965.17   -6.606   8  -7.099   8   ----   0   ----   0   ----   0   ----   0
1966.18   -6.610  13  -7.094  13   ----   0   ----   0   ----   0   ----   0
1991.48   -6.626  14  -7.178  49   ----   0   ----   0   ----   0   ----   0
1992.50   -6.584  20  -7.162  29   ----   0   ----   0   ----   0   ----   0
1993.51   -6.691  34  -7.153  34  -6.995  19  -5.749  19   ----   0   ----   0
1994.52   -6.667  15  -7.156  16  -6.962  13  -5.712  13   ----   0   ----   0
1995.53   -6.619   7  -7.175   8  -6.997   7  -5.684   7   ----   0   ----   0
1996.54    ----    0  -7.120  12   ----   0   ----   0   ----   0   ----   0
1997.56    ----    0  -7.148  53   ----   0   ----   0   ----   0   ----   0
1998.57   -6.657  10  -7.156  41   ----   0   ----   0   ----   0   ----   0
1999.58   -6.647  20  -7.157  30  -6.890   1  -5.600   1   ----   0   ----   0
2000.59   -6.560   1  -7.143  12  -6.821   9  -5.611  19   ----   0   ----   0
2001.60   -6.604   5  -7.131  74  -6.930   1   ----   0   ----   0   ----   0
2002.62   -6.628   8  -7.128  34  -6.747   7  -5.553   3   ----   0   ----   0
2003.63   -6.595  15  -7.123  20   ----   0   ----   0   ----   0   ----   0
```

```
2004.64  -6.618  5  -7.111 10  -6.678 4  ----    0  ----    0  ----    0
2005.65  -6.659 11  -7.120 22  ----   0  ----    0  ----    0  ----    0
2006.66  -6.596  5  -7.113 45  -6.680 1  ----    0  ----    0  ----    0
2007.68  -6.645 17  -7.124 32  -6.707 3  ----    0  ----    0  ----    0
2008.69  -6.658  7  -7.118 58  ----   0  ----    0  ----    0  ----    0
2009.70  -6.650  2  -7.150  4  ----   0  ----    0  ----    0  ----    0
2010.71  -6.555 26  -7.134 27  -6.775 4  -5.613  3  ----    0  ----    0
2011.72  -6.576 17  -7.141 18  -6.814 5  -5.697  4  ----    0  ----    0
2012.74  -6.645 22  -7.116 24  -6.812 6  -5.628  5  ----    0  ----    0
2013.75  -6.651 38  -7.125 41  -6.879 3  -5.679  3  ----    0  ----    0
2014.76  -6.630 40  -7.134 40  -6.928 3  -5.724  3  -6.996  6  -6.080  6
```

4. Brightness variations in the B and V bands

The B- and V-band observations span the years from 1953 through 2015 and include two equinoxes. The brightness changes during this time interval are shown in Figs. 4 and 5. Dimming is evident near the 1966 and 2007 equinoxes (MJDs 39300 and 54300, respectively).

We computed a value called the *sub-latitude,* which is the average of the absolute values of the planetographic sub-Earth and sub-Sun latitudes, for the time of each observed magnitude. Since Uranus orbits at a distance of about 20 AU the sub-latitude, sub-Earth latitude and sub-Sun latitude are nearly equal.

The relationships between magnitudes and sub-latitude are shown in Figs. 6 and 7, and they are quantified in Table 6. The change of magnitude per degree of latitude for the northern and southern hemispheres combined (black lines in the figures) is $-0.48 \times 10^{-3}$ in the B-band and $-0.84 \times 10^{-3}$ in the V-band.

When the data are separated by the hemisphere of the sub-latitude (orange and green symbols and lines in Figs. 6 and 7) they imply the planet was brighter when southern latitudes were centered on the disk in both bands.

The oblate figure of Uranus causes the apparent surface area of its disk to increase along with sub-latitude. The excess over the area projected at sub-latitude zero is proportional to sine-squared of the sub-latitude. Since the ratio of the planet's equatorial radius to its polar radius is 1.0235, the excess is 0.0235 at 90 degrees. Fig. 8 shows the observed excess flux of Uranus plotted versus the sine-squared term in the B-band and V-bands along with best-fitting lines in color. The black lines indicate the geometrical relationship between surface area and oblateness. The observed slope for the V-band (0.0575 +/- 0.0030) exceeds the slope for oblateness (0.0235) by more than 10 standard deviations. However, the observed slope for the B-band (0.0330 +/- 0.0070) only exceeds the oblateness slope by about 1.3 standard deviations. Thus our results for the B- and V-bands are in general agreement with those of Lockwood and Jerzykiewicz (2006) for the b- and y-bands.

Table 6. Magnitude as a function of sub-latitude

```
Band    Zero Point    +/-       Slope*      +/-
 B        -6.61       0.02      -0.48       0.11
 V        -7.11       0.02      -0.84       0.04
 R        -6.69       0.02      -5.33       0.30
 I        -5.57       0.02      -2.79       0.41
```
* Milli-magnitudes per degree of latitude

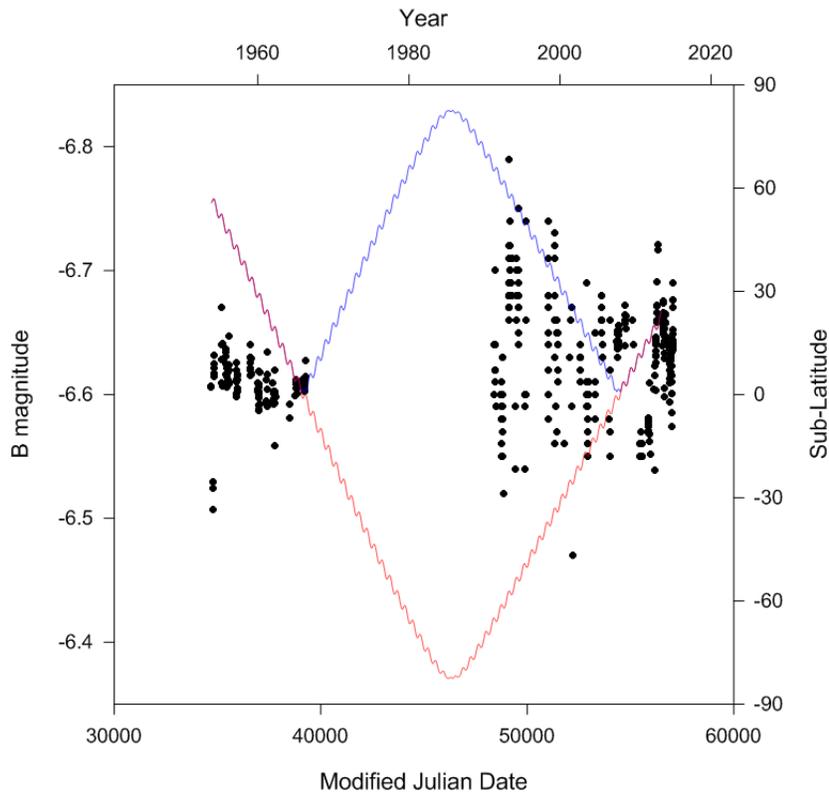

Fig. 4. Brightness variation of Uranus in the B-band between 1953 and 2015. The early data are from Serkowski (1961) and Jerzykiewicz and Serkowski (1966). The uncertainties for the early data are about 0.01 magnitude and for the later data they are a few hundredths of a magnitude. The red line represents the sub-latitude and the blue line is the absolute value of the sub-latitude.

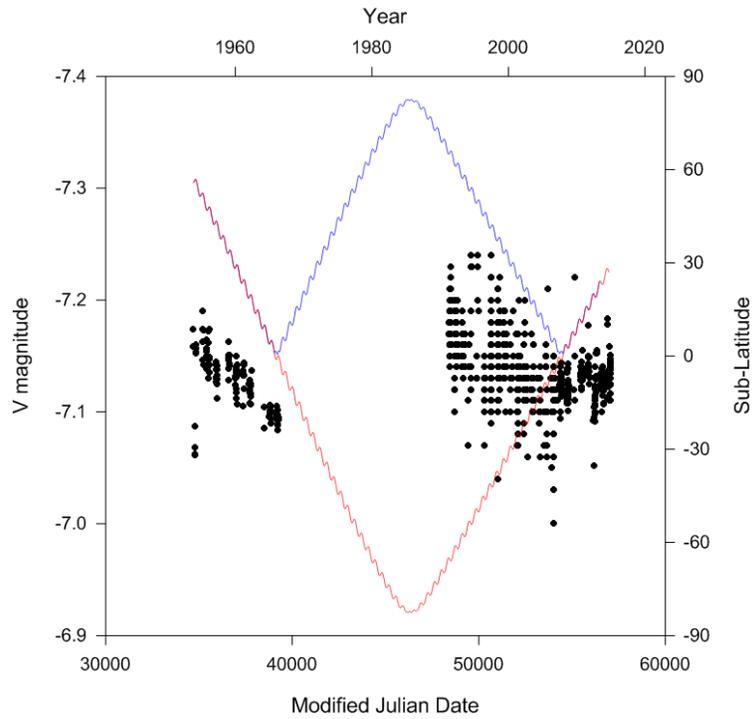

Fig. 5. Same as Fig.4 but for the V-band.

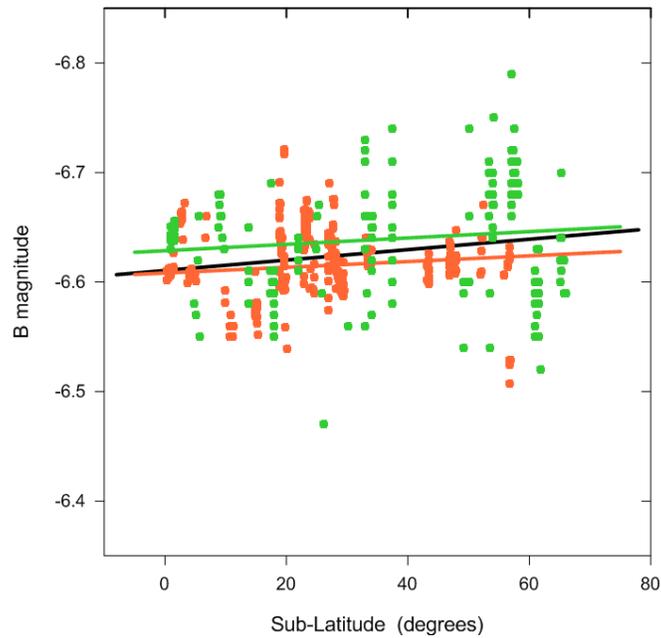

Fig. 6. B-band magnitude as a function of sub-latitude. The black line is the best fit to all the data . The orange and green data and lines are for data where the sub-latitude was north and south of the equator, respectively.

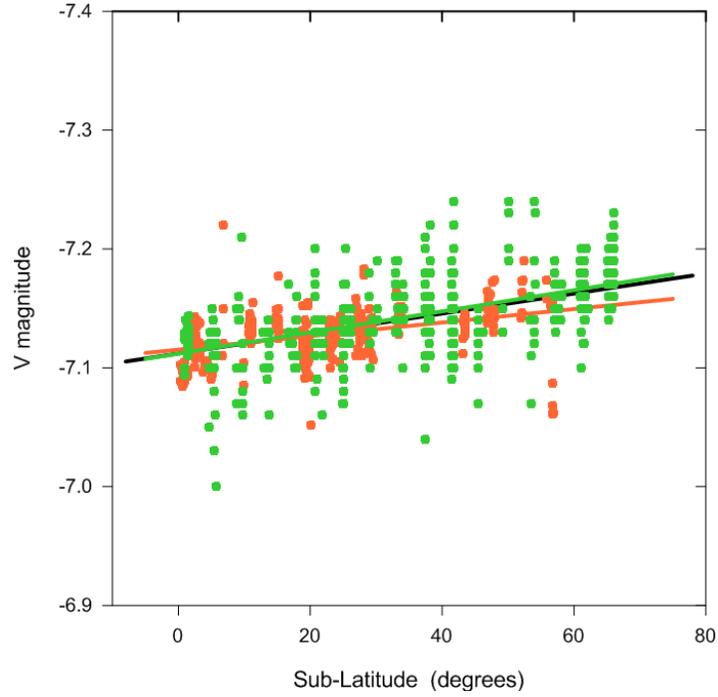

Fig. 7. Same as Fig. 6 but for V-band magnitude.

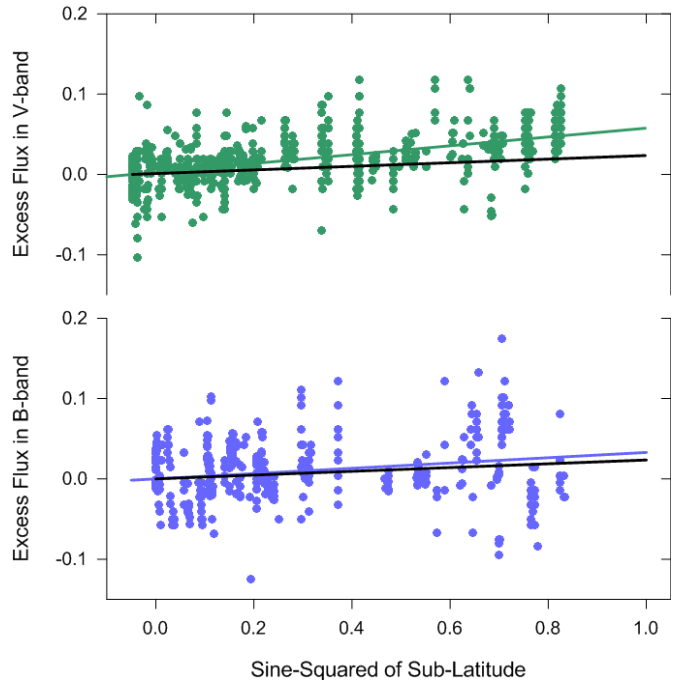

Fig. 8. The excess of flux relative to that at sub-latitude zero is plotted versus the sine-squared of sub-latitude for the B- and V-bands. The black lines represent the corresponding excess of projected surface area due to planetary oblateness.

5. Large brightness variations in the R and I bands

The R- and I-band observations span the years from 1993 through 2014 and include the equinox of 2007. The brightness variations illustrated in Figs. 9 and 10 are much larger than those in B and V. The rate of change of magnitude with sub-latitude is $-5.33 \times 10^{-3}$ magnitudes per degree for the R-band and $-2.79 \times 10^{-3}$ for the I-band, and they correspond with brightness variations measured in tenths of magnitudes (tens of percents).

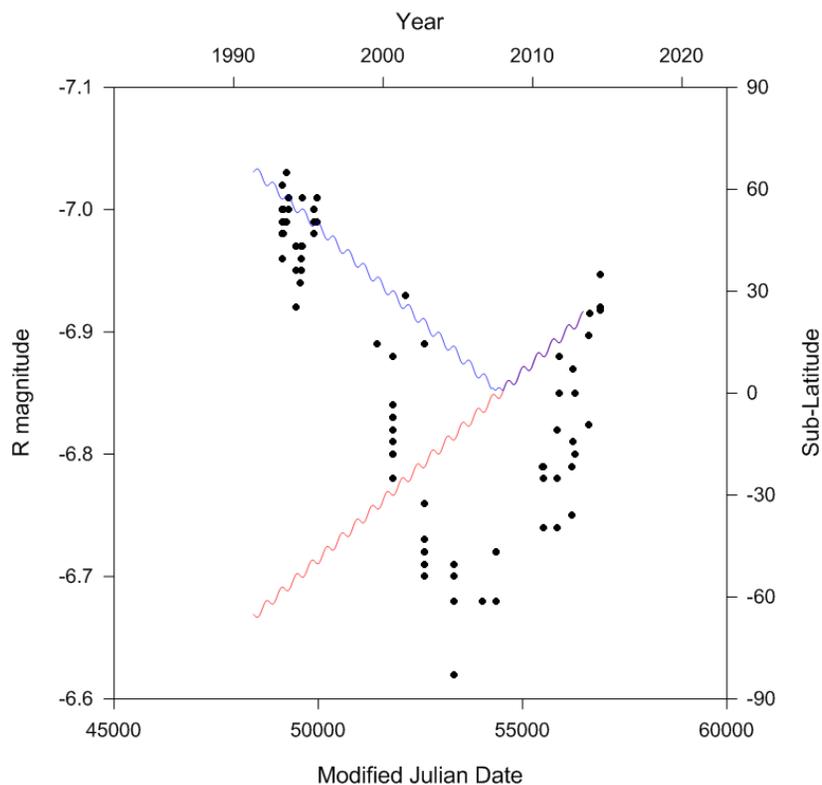

Fig. 9. Brightness variation of Uranus in the R-band between 1993 and 2015. The uncertainties are a few hundredths of a magnitude. The red line represents the sub-latitude and the blue line is the absolute value of the sub-latitude.

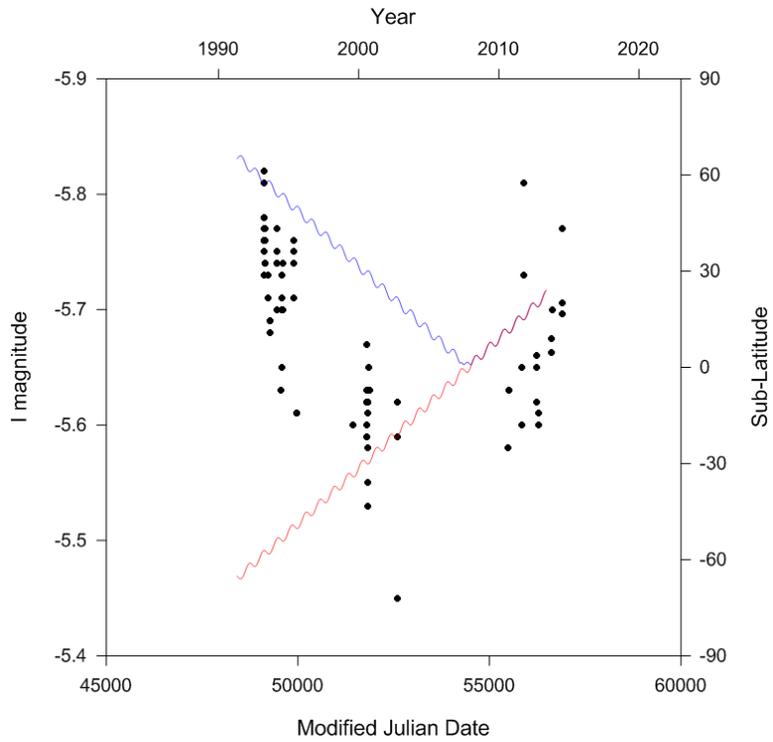

Fig. 10. Same as Fig. 9 but for the I-band.

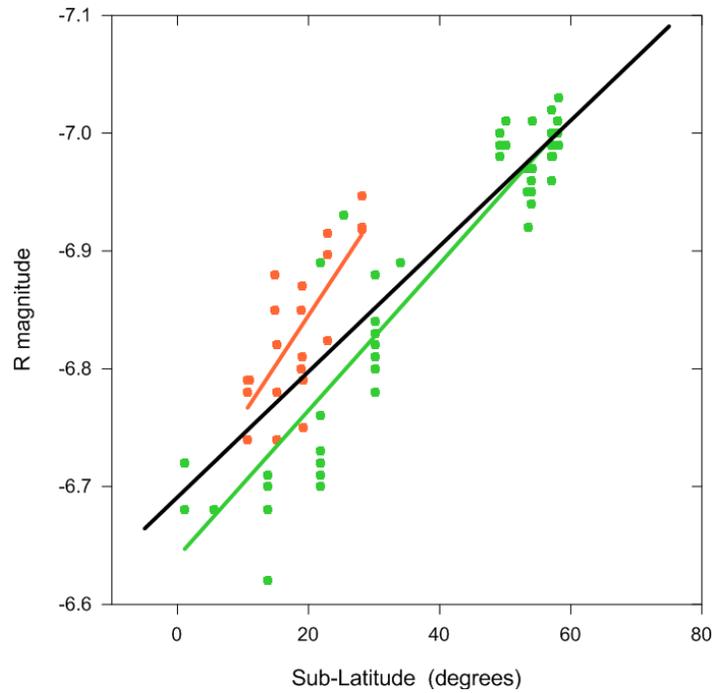

Fig. 11. R-band magnitude as a function of sub-latitude. The black line is the best fit to all the data . The orange and green data and lines are for data where the sub-latitude was north and south of the equator, respectively.

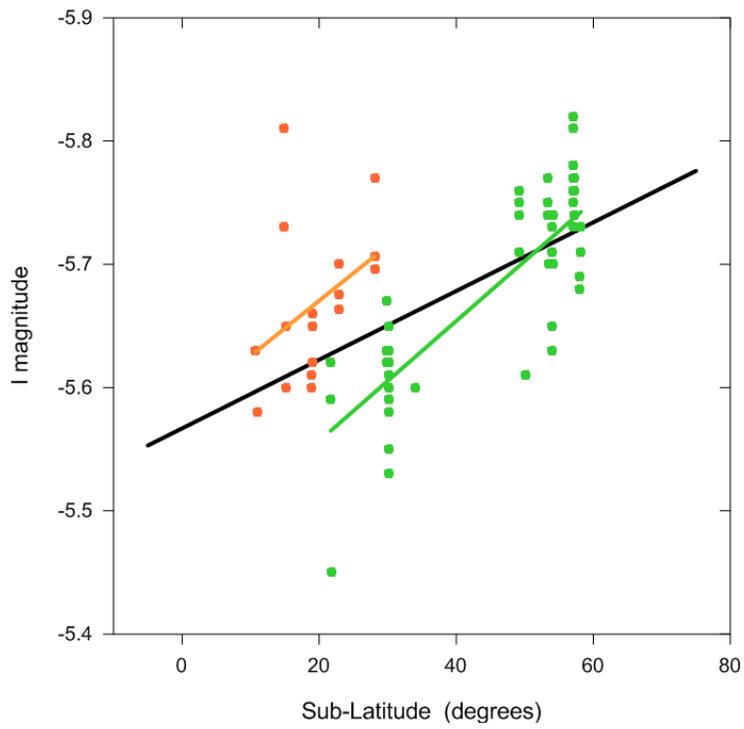

Fig. 12. Same as Fig. 11 but for I-band magnitude.

6. Brightness contributions from the Uranian satellites, rings and clouds

There are several additional factors which could cause observed brightness changes. Karkoschka (1997) reports albedos for the largest Moons and rings over the 340-910 nm range. In the V filter, Titania is ~2500 times dimmer than Uranus (Astronomical Almanac, 2013). The albedo of Titania divided by that of Uranus is ~0.5 for the V filter but is ~5 for the I filter based the albedos in Karkoschka (1997, 1998). This is a factor of 10 increase and, hence, Titania will still be ~250 times dimmer than Uranus in the I filter. Therefore, Titania will create a negligible ~0.004 magnitude increase when it enters the photometer field of view for the I filter. In rare cases when the phase angle is below 0.1 degree this may rise to 0.01 magnitudes.

The rings are 2000 times fainter than Uranus in visible light (Karkoschka, 1997). The situation is a little different for the I filter. Calculations utilizing the ring dimensions in French et al. (1987) along with the albedos in Karkoschka (2001a) suggest that in the I filter the rings will change the brightness of the Uranus system by ~0.01 magnitude in the I filter.

A third factor, clouds, may create a diurnal brightness change. Karkoschka (2001b) measured the brightness of several clouds at wavelengths between 0.66 and 2.03 µm. Even near equinox when the brighter Northern Hemisphere clouds are less foreshortened, it is unlikely that diurnal variations should exceed 0.01 magnitudes in this study unless there is an unusually large cloud.

7. The polar regions, methane depletion and hemispheric asymmetry

The observed disk-integrated variations of magnitude with latitude are consistent with the characteristic brightness associated with the North and South Polar Regions (NPR and SPR). As early as the 1970s, there was evidence that the planet's SPR was brighter than the surrounding areas in red and near-infrared wavelengths (Price and Franz, 1978 and Radick and Tetley, 1979). Furthermore, Price and Franz report that the polar brightening was much more distinct in 750 nm compared to 619 nm. Twenty years later, HST images show this same trend (Rages et al, 2004). Rages and co-workers also present images taken at 619 nm which show only small brightness changes taking place in the SPR. Images taken in the 890 methane band show little albedo change across the disk (Karkoschka, 1998). One explanation which Rages et al. (2004) propose is the development of clouds of condensed methane over the SPR which later dissipate leaving behind a darker area. This mechanism may also be taking place in the NPR where brightening since 2007 is consistent with higher reflectivity in the northern hemisphere. Infrared images also show bright belts in the Northern Hemisphere (Schmude, 2014).

Methane depletion in the planet's polar regions, as reported by Sromovsky et al (2014), may be associated with the disk-integrated brightness variations we observed as well as the disk-resolved polar brightness discussed above. The weakening of methane absorption bands (which dominate the red and near-IR spectrum) at high latitudes should result in greater reflectivity toward the poles as seen in the R- and I-band results. Conversely, the absence of strong methane bands in the blue and green portions of the spectrum are consistent with the smaller brightness variations recorded in the B and V.

Finally, the observed disparity between brightness in the northern and southern hemispheres agrees fairly well with the asymmetry findings of Karkoschka (2001b). The bottom portion of Karkoscka's Fig. 4 indicates that the southern hemisphere reflects a little more strongly at 430 nm as does our B-band result (Fig. 6). Meanwhile, he finds that the northern hemisphere reflects much more strongly beyond 600 nm which agrees with our R- and I-band results (Figs. 11 and 12). There is a minor disagreement, however, at 550 nm where Karkoscka indicates a slightly brighter northern hemisphere while the V-band data suggests a slightly brighter south (Fig. 7).

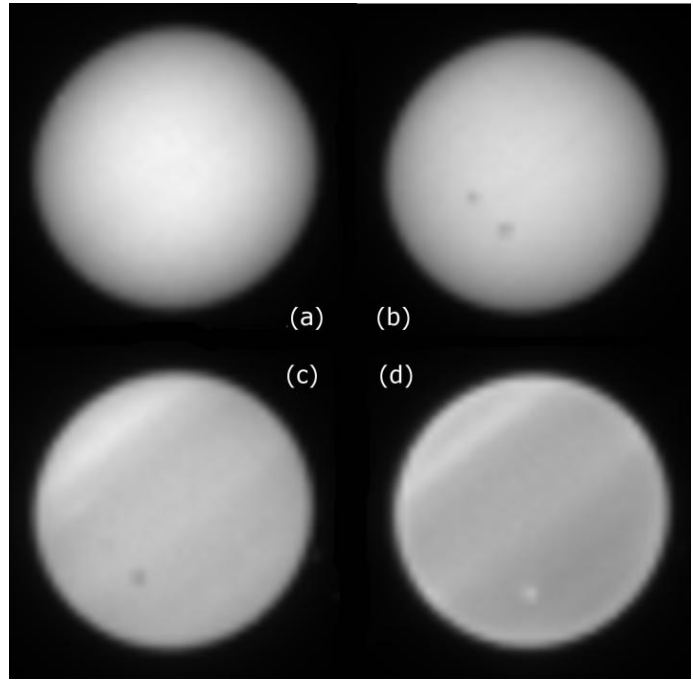

Fig. 13. Uranus is a relatively featureless disk at blue (a) and green (b) wavelengths while the polar regions can be brighter than equatorial latitudes at red (c) and near-IR (d) wavelengths. These HST images were recorded near the time of Uranian equinox on 2007 July 29. The latitude limits of the polar regions have been measured at 44 +/- 3 degrees north by Schmude (2014 and 2015a) and 50 +/- 0.5 degrees south by Smith 1987). The image names and filters are as follows: a, U9Z70702M, F467M; b, U9Z70503M, F547M; c, U9Z70505M, F658N; and d, U9Z70608M, F850LP.

8. Reference magnitudes, albedos and colors

We adopt the brightness at equinox (zero point numbers in Table 6) as the B, V, R and I reference magnitudes for Uranus. U filter data was also recorded in 2014 by one of us (RWS) when the sub-latitude was +27.0. By applying the slope of 'magnitude versus sub-latitude' for the B-band in Table 6, we have arrived at the reference value in Table 7.

Additionally, one of the authors (REB) obtained $R_C$ and $I_C$ magnitudes on the Johnson-Cousins system in 2014. We applied the slope of 'magnitude versus sub-latitude' for the R-band in Table 6 to the Rc values and, likewise, applied the I-band slope to the Ic values. The averages of the resulting equinoctial magnitudes were taken as the reference magnitudes in $R_C$ and $I_C$. These magnitudes are being reported here for the first time.

All reference values are listed in Table 7 along with implied magnitudes at sub-latitude 90 degrees. For comparison, the table also lists U, B and V magnitudes from Karkoschka (1998) and from the Astronomical Almanac. Karkoschka's B and V magnitudes, which were obtained in July 1995 when the sub-latitude of Uranus was 49 degrees, are toward the bright end of the range from this paper as would be expected given the relatively high latitude. The B and V values from the Almanac lie within the values for 0 – 90 degree sub-latitude from this paper. The U magnitudes from both Karkoschka and from the Almanac are slighter brighter than the range from this paper but there are within the uncertainty.

Table 7. Magnitude summary

|  | U | B | V | R | $R_C$ | I | $I_C$ |
|---|---|---|---|---|---|---|---|
| Reference* | −6.28 | −6.61 | −7.11 | −6.69 | −6.84 | −5.57 | −6.00 |
| Uncertainty* | 0.05 | 0.02 | 0.02 | 0.02 | 0.05 | 0.02 | 0.05 |
| 90 degrees* | −6.32 | −6.65 | −7.19 | −7.17 | −7.32 | −5.83 | −6.25 |
| Uncertainty* | 0.05 | 0.02 | 0.02 | 0.03 | 0.05 | 0.04 | 0.05 |
| Almanac** | -6.35 | -6.63 | -7.19 | --- | --- | --- | --- |
| Karkoschka*** | -6.36 | -6.64 | -7.17 | --- | --- | --- | --- |

* This paper

** Astronomical Almanac (2013)

*** Karkoschka (1998) value corresponds to a solar phase angle of 0.7 degree

The geometric albedo is defined as the ratio of the observed flux for a planet to that of a perfectly reflecting Lambertian disk. Taking the V filter as an example, the magnitude of the Sun is –26.75 (Table 8), so the ratio of the luminosity of Uranus (magnitude –7.11 from Table 7) to that of the Sun is $1.39 \times 10^{-8}$ as shown in Equation 3.

$$\text{Lratio} = 10^{\{(-26.75+7.11)/2.5\}} = 1.39 \times 10^{-8}$$

(3)

The average disk radius seen at equinox, $r$, is 25264 km (including oblateness). The astronomical unit (AU) distance is $149.6 \times 10^6$ km, hence there is an area factor, $\sin^2(r/AU)$ or $2.85 \times 10^{-8}$. Therefore the geometric albedo, p, is 0.488 as shown in Equation 4.

$$p = \text{Lratio}/ \sin^2 (r/AU) = 0.488$$

(4)

Table 8. Solar magnitudes

```
      U*       B*       V*       R*       Rc**     I*       Ic**
    -25.90   -26.10   -26.75   -27.29   -27.15   -27.63   -27.49
     * Livingston (2001)
    ** Binney and Merrifield (1998).
```

The albedo values in Table 9 range from a maximum of 0.561 in the B-band to a minimum of 0.053 in the I-band. This variation is also reflected in the Uranian colors which are described next.

Table 9. Geometric albedos

|  | U | B | V | R | $R_C$ | I | $I_C$ |
|---|---|---|---|---|---|---|---|
| Value | 0.500 | 0.561 | 0.488 | 0.202 | 0.263 | 0.053 | 0.089 |
| Uncertainty | 0.025 | 0.011 | 0.010 | 0.004 | 0.013 | 0.001 | 0.004 |

The reference colors listed in Table 10 reveal the peculiar spectral energy distribution of Uranus. The planet is moderately red according to the U-B and B-V indices but quite blue according to the indices at longer wavelengths. These unusual colors are due to the strong methane bands which blanket the red and near-IR portions of the spectrum. Table 10 also lists the colors at 60 degrees sub-latitude as inferred from slopes of magnitude versus sub-latitude in Table 1. The B-V and V-R values grow larger which indicates general reddening in the blue-through-red portion of the spectrum. However, the R-I index decreases because the slope of R magnitudes is greater than that of I magnitudes.

Table 10. Colors

|  | U-B | B-V | V-R | V-Rc | R-I | Rc-Ic |
|---|---|---|---|---|---|---|
| Reference | +0.32 | +0.50 | -0.42 | -0.27 | -1.12 | -0.84 |
| 60 degrees | --- | +0.52 | -0.15 | --- | -1.27 | --- |
| Uncertainties | 0.05 | 0.02 | 0.02 | 0.05 | 0.02 | 0.05 |
| Solar values | +0.20 | +0.65 | +0.54 | +0.40 | +0.34 | +0.34 |

9. Conclusion

We have reported on brightness changes of Uranus and interpreted them in terms of the planet's geometrical and geophysical properties. Variations in the B- and V-bands were found to be small and in generally good agreement with b- and y-band results reported by Lockwood and Jerzykiewicz (2006). Variations in the R- and I-bands, reported here for the first time, were quite large and are quite interesting. Continued observations, especially at red and near-IR wavelengths are warranted. We also list Uranian reference magnitudes, colors and albedos.

Acknowledgments

RWS thanks the library staff at Gordon State College for their assistance in locating source material used in this study. The data used in Figure 13 was retrieved from the MAST archive of Hubble Space Telescope data at the STScI.